\documentclass[letter,10pt]{article}

\usepackage{amsmath, amssymb}
\usepackage{amsthm}
\usepackage{algorithm}
\usepackage{algorithmic}
\usepackage{color, array, colortbl}
\usepackage{graphicx}
\usepackage{multirow}
\usepackage{color, array, colortbl}
\usepackage{graphicx}
\usepackage{comment}
\usepackage{subfig}
\usepackage{multirow}
\usepackage{framed,color}
\usepackage{color, array, colortbl}
\definecolor{shadecolor}{rgb}{0.9,0.9,0.9}
\newcommand{\mathsym}[1]{{}}

\newcommand{\cancel}[1]{}

\usepackage{amsmath}

\usepackage{graphicx}

\usepackage{epsfig}
\usepackage[german,english]{babel}
\usepackage{fullpage}

\date{}

\title{A Federated CloudNet Architecture:\\The PIP and the VNP Role}

\author{
Ernesto Abarca \quad Johannes Grassler\\Gregor Schaffrath \quad Stefan Schmid\\
\small{T-Labs \& TU Berlin}
}

\begin{document}

\maketitle

\begin{abstract}
 We present a generic and flexible architecture to realize \emph{CloudNets}: virtual networks connecting cloud resources with resource guarantees. Our architecture is federated and supports different (and maybe even competing) economical roles, by providing explicit negotiation and provisioning interfaces. Contract-based interactions and a resource description language that allows for aggregation and abstraction, preserve the different roles' autonomy without sacrificing flexibility. Moreover, since our CloudNet architecture is \emph{plugin based}, essentially all cloud operating systems (e.g., OpenStack) or link technologies (e.g., VLANs, OpenFlow, VPLS) can be used within the framework.

 This paper describes two roles in more detail: The \emph{Physical Infrastructure Providers (PIP)} which own the substrate network and resources, and the \emph{Virtual Network Providers (VNP)} which can act as resource and CloudNet brokers and resellers. Both roles are fully implemented in our wide-area prototype that spans remote sites and resources.
\end{abstract}

\section{Introduction}

The virtualization paradigm is arguably the main innovation motor in today's Internet.
Especially node virtualization revolutionized the server business over the last years, and in today's cloud resources are fully virtualized. 
However, node virtualization alone is meaningless without access to the cloud resources. Thus, to provide performance guarantees, cloud virtualization needs to be extended to the \emph{communication network}.

The concept of \emph{CloudNets} (short for \emph{cloud networks}) takes the virtualization paradigm one step further, envisioning a unified approach which combines node and link virtualization and offers Quality-of-Service (QoS) guarantees, both on the nodes and the links. Basically, a CloudNet describes a virtual network topology where the virtual nodes represent cloud resources (e.g., storage or computation) which are connected by virtual links.

While CloudNet concepts are already emerging in the context of data centers (e.g.,~\cite{netperf,d3}), we expect that in the future, Internet Service Providers (ISP) will also offer flexibly specifiable and on-demand virtual networks, connecting (heterogeneous) cloud resources with connectivity guarantees. Such an \emph{elastic} wide-area network (WAN) connectivity is attractive in many settings. For example, for inter-site data transfers or state synchronization of a distributed application (such as online gaming). Other use cases are the \emph{spill-over} (or \emph{out-sourcing}) to the public cloud in times of resource shortage in the private data center, or the distribution of content to CDN caches. 

Note that a CloudNet may not even specify the locations of its constituting resources, and the mapping of the CloudNet can hence be \emph{subject to optimization}. In fact, the resources of the CloudNets can be \emph{migrated} over time. For example, latency-critical CloudNets (e.g., realizing a game, an SAP or a social networking service) can be dynamically migrated closer to the users,
 while delay-tolerant CloudNets (e.g., for large-scale computations or bulk data storage) are run on the remaining servers. 
 Moreover, resources allocated to a CloudNet can be scaled up or down depending on the demand at the different sites. Finally, the decoupling of the CloudNet from the underlying physical infrastructure can also improve reliability, as networks can seamlessly switch to alternative cloud and link resources after a failure, or for maintenance purposes (see e.g., the Amazon outage in April 2011).

This paper presents the anatomy and prototype implementation of a flexible and federated CloudNet architecture. In particular, we will describe the Physical Infrastructure Provider (PIP) role and the Virtual Network Provider (VNP) role. The two roles communicate requirements and allocations via clear negotiation interfaces and using a generic resource description language. Since interfaces between players are generic and since we do not distinguish between physical and virtual resources, a recursive role concept is supported (e.g., a VNP sub-structured into other VNPs).

Moreover, a high generality is achieved by a \emph{plugin architecture} which allows for replacement of underlying technologies and operating systems. 

\section{Background}\label{sec:arch}

Our architecture is based on the following visions and concepts.

\subsection{The Road to CloudNets}

Inter-ISP QoS connectivity has already been discussed for many years and is still not widely supported, so why should wide-area and multi-provider CloudNets become a reality now? We believe that the answer lies in the economical incentives, and the recent virtualization and \emph{Software-Defined Networking (SDN)} trends (e.g.,~\cite{j-sdn}). Content and service providers (e.g., Netflix, Amazon, etc.) as well as content distribution network providers (e.g., Akamai) have become powerful players in the Internet and have stringent resource requirements. CloudNets can be one way for an ISP to monetarize its infrastructure by offering flexible service deployments.
For example, in the architecture described in this paper, physical infrastructure providers play a central role in the CloudNet contract hierarchy and participate in the service negotiation, which allows them to become a service partner \emph{at eye-level} (rather than being a pure bitpipe provider).

We expect that CloudNets will first be deployed inside a single ISP. Such an innovative ISP may benefit from a more efficient resource management,\footnote{Google is an example of a recent unilateral initiative in the context of SDN.} and may have a first-mover advantage by offering new flexible and elastic services.  In the context of an ISP, especially the possibility to deploy (and migrate) services closer to the eyeballs of the users may increase the productivity and revenues: it is known that a better web performance directly translates into higher productivity at \emph{Google}, and higher revenue at \emph{Microsoft Bing} or \emph{Amazon}.

Over time, it can make sense for so far independent CloudNet providers to \emph{collaborate}. For example, two providers with a local footprint can offer more global CloudNets, or providers can flexibly lease resources to each other to compensate for time-of-day dependent resource shortages. 

\subsection{Economical Roles}

The advent of CloudNets may create a tussle among service and infrastructure providers over who should operate and who should manage the corresponding networks. In~\cite{visa09virtu}, we identified two additional roles (i.e., potential market \emph{players}) besides \emph{Physical Infrastructure Providers (PIPs)} and \emph{Service Providers (SPs)}: \emph{Virtual Network Providers (VNPs)} (essentially ``resource brokers'') for assembling virtual resources from one or multiple PIPs into a virtual network, and \emph{Virtual Network Operators (VNOs)} for the installation and operation of the CloudNet provided by the VNP according to the needs of the SP.
Our CloudNet architecture defines standardized interfaces between the players to automate the setup of virtual networks (by using a common control plane). In this paper, we will focus on the PIP and VNP roles: both roles have already been fully implemented in our prototype.

\subsection{Specification and Specificity}\label{sec:flerd}

A powerful \emph{Resource and CloudNet Description
Language (RDL)}~\cite{icccn12stefan} (a.k.a.~FleRD) is used in our architecture. It revolves around basic \emph{NetworkElement (NE)}
objects (for both nodes \emph{and links}!) that are interconnected via
\emph{NetworkInterfaces (NI)} objects.  Keeping these
objects generic has the side effect that descriptions of resource aggregations,
or non-standard entities (e.g., clusters of providers) is trivially
supported:
they may be modeled as \emph{NEs} of an appropriate type and included as
topological elements. Thus, we can for example also describe mappings in the
context of a reseller.
Concretely, \emph{NE} properties are represented as a set of \emph{attribute-value pair}
objects labeled as \emph{Resource} and \emph{Features}. The meaning of
resources here is canonic and resources may be shared amongst \emph{NEs}.
Features represent any type of property that is not a resource (e.g., CPU
architecture). 

\begin{figure*}[ht]
\centering
\includegraphics[width=.45\textwidth]{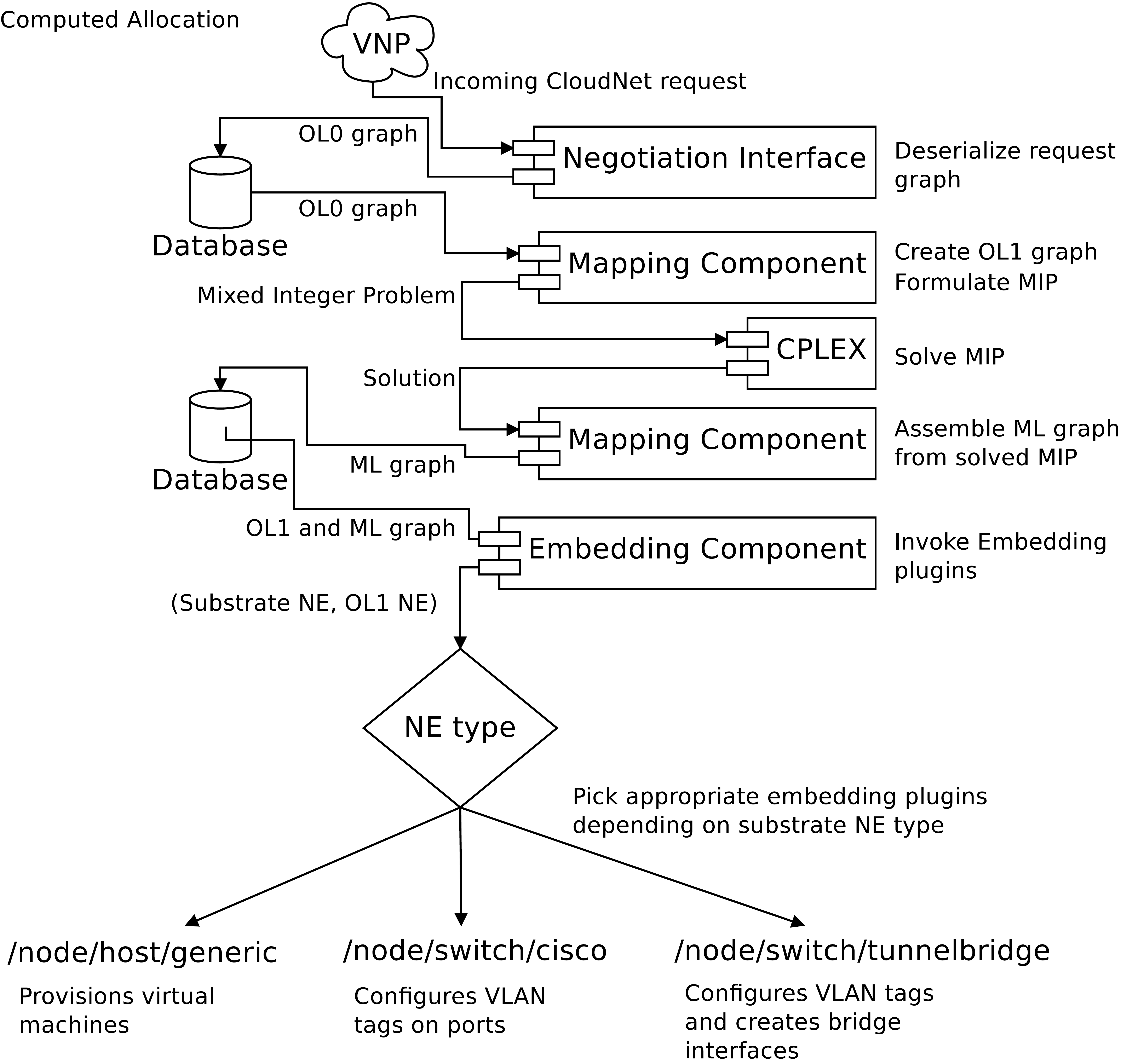}~~~~~~~~
\includegraphics[width=.37\textwidth]{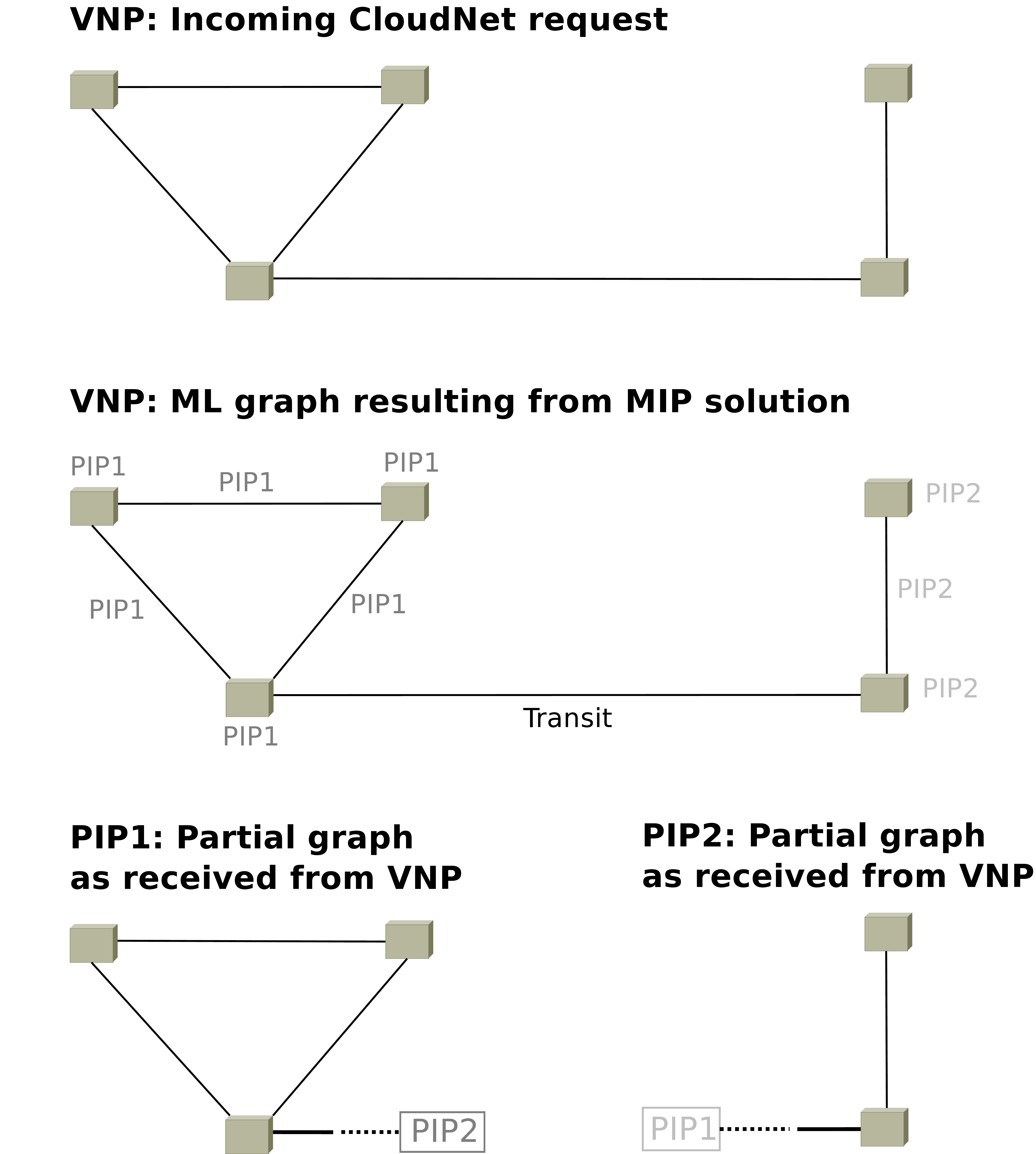}\\
\caption{\emph{Left:} The CloudNets architecture as implemented by the PIP role. \emph{Right:} Stages of a CloudNet request from VNP to PIP.}
\label{fig:architecture}
\end{figure*}

One challenge in the communication of specifications and RDL elements across different roles is to ensure consistency, especially for non-topological requirements. For instance, imagine a service provider specification only requires that two nodes are \emph{binary compatible} (e.g., both are 32 bit architectures or both are 64bit architectures). How can a VNP in charge of distributing such a CloudNet across multiple PIPs fulfill the specification? A simple solution would be that the VNP chooses one of the different options (e.g., both 64 bit) before forwarding the request to the different PIPs. However, this solution comes with a major disadvantage, namely an unnecessary specificity and hence a loss in flexibility: as the VNP has a limited view on the infrastructure of the PIPs only, the VNP's choice may lead to inefficient allocations (e.g., the two nodes are mapped unnecessarily far away). Alternatively, the VNP may assign the corresponding NEs unique IDs which can be used by the PIPs to agree on a choice themselves. The required communication to achieve this may go via the VNP again, or occur directly between the PIPs. In the following, we will not discuss these options in more detail. 

\subsection{Plugins}

Our architecture is plugin-based to facilitate extensions and
adaptations to new networking and virtualization technologies. Plugin-based
operation hinges on a feature of the resource description language, namely \emph{typed
NetworkElements}:
The \emph{NetworkElement}  objects all have a hierarchical \emph{attribute}
field (this also applies to the \emph{NetworkInterface}, \emph{Resource} and
\emph{Feature} objects). This field is a string of hierarchy levels, like a path in a Unix file system (e.g.,
\texttt{/node/host/generic}).

Much like a CloudNet, any player's substrate can be described in terms of the
RDL. Hence all substrate NEs and all CloudNet NEs will have an \emph{attribute}
field. The higher order hierarchy levels of a CloudNet's NEs type will be used
in the course of the mapping process to determine which substrate NEs they can
be allocated to (i.e., both \texttt{/node/host/mainframe} and
\texttt{/node/host/server} might be suitable allocations for a
\texttt{/node/host/generic} CloudNet NE).

Once a CloudNet's resources have been mapped to a player's substrate resources,
the lower levels of the substrate NE's type will be used to
determine which plugin to use to provision the CloudNet NEs in question.
Depending on the substrate NetworkElement's hardware or software (e.g.,
Layer 2 VPNs, Ethernet, OpenFlow, MPLS, mainframes, VM hosts), the appropriate embedding
plugin can be chosen to embed all the CloudNet's resources mapped to the
substrate element in question.

\section{Testbed \& Prototype}\label{sec:prototype}

The CloudNets prototype is a proof-of-concept implementation of the architecture described above.
Currently, the prototype implements the PIP
business role (the infrastructure provider) as well as the VNP business role
(the resource broker). As a link virtualization plugin we use \emph{tagged VLANs}: each virtual
link is realized as a VLAN. The VLANs provide isolation and enable the 
demultiplexing of frames to the corresponding virtual machines (to which 
the connection looks like an Ethernet link).

In the following, we will first give an overview of the testbed and then
discuss the prototype. In the context of the prototype, we will refer to
virtual nodes as \emph{VNodes} and to virtual links as \emph{VLinks}. Moreover,
we will sometimes distinguish between the \emph{mapping} of a CloudNet (i.e., where to
allocate the CloudNet in the substrate) and the \emph{embedding} of a CloudNet
(the actual implementation of said allocation).

\subsection{Testbed}

\begin{figure*}[ht]
\centering
\includegraphics[width=.65\textwidth]{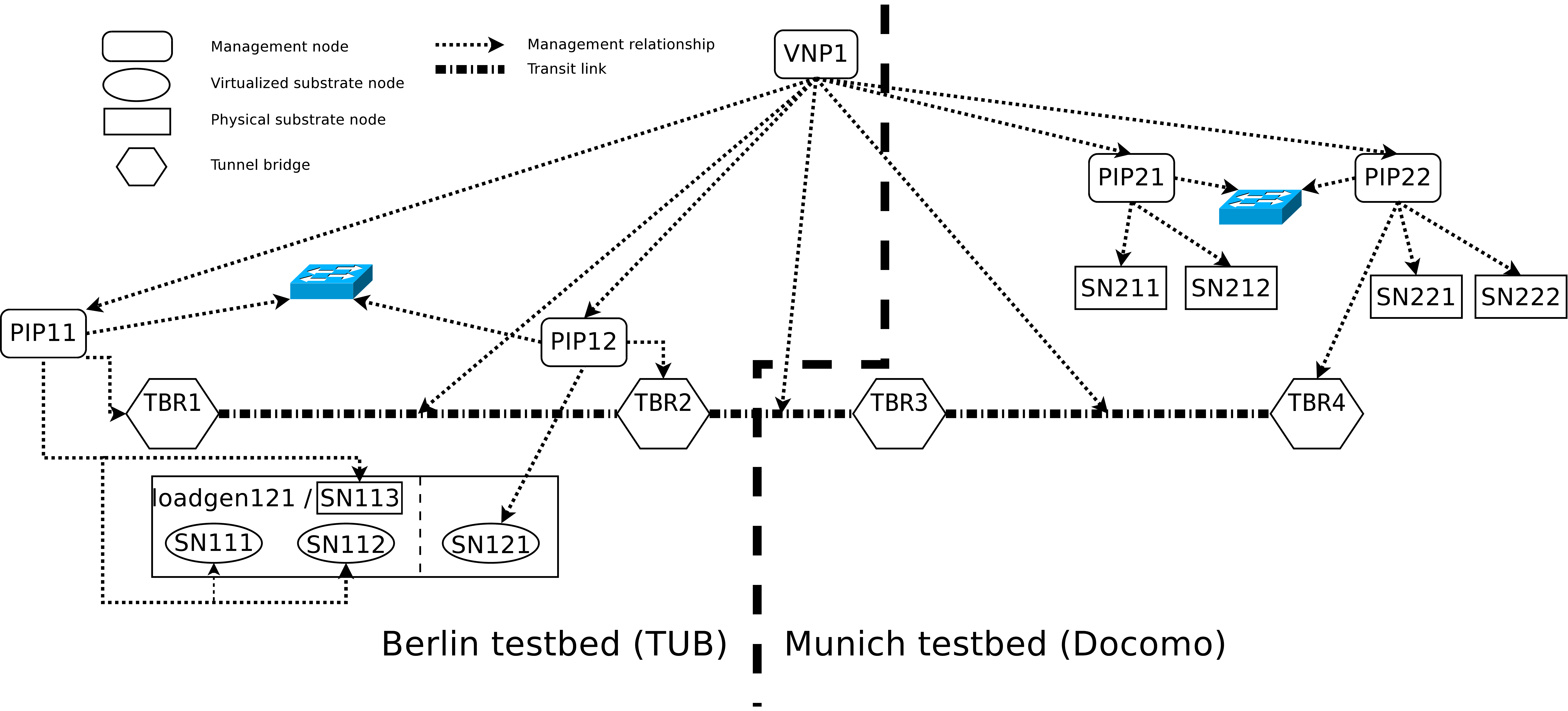}\\
\caption{Overview diagram of a test topology for VNP level operations, spanning
both the TU Berlin and Docomo testbeds. Note the physical substrate node
SN113 (loadgen121) that is also serving as a host for virtualized substrate
nodes.}
\label{fig:testbed}
\end{figure*}

Our network virtualization prototype is run on two separate testbed
environments (see Figure~\ref{fig:testbed} for an overview of the available
substrate nodes), one at TU Berlin (the \emph{Routerlab}) and
one at NTT DoCoMo Eurolabs, Munich. Both testbeds have a Cisco 4500 series
switch carrying both virtual network \emph{data plane} VLANs and \emph{testbed
management} VLANs. Moreover, both testbeds consist of Sun X4150 machines hosting both
substrate resources and the virtual machines running the various roles'
management software. Since there are not enough physical nodes available in the
\emph{Routerlab}, some substrate nodes are KVM virtual machines hosted on one
physical node. In the testbed in Munich there are more physical machines
available, hence every substrate node is a dedicated physical machine, with the
provider management nodes being virtual machines hosted on one physical
machine. Substrate nodes, management nodes and virtual nodes run Linux
(\texttt{Ubuntu 8.04}). Virtual nodes can either be fully virtualized
\emph{KVM} virtual machines where physical substrate nodes are available, or
paravirtualized \emph{Xen} virtual machines.

The substrate resource allocation for the CloudNet embeddings (CPU and memory
resources on virtual machine hosts, and network links for connections) is
computed with an optimized \emph{Mixed Integer Program (MIP)}, see Section~\ref{ssec:mip}. Once the
resource allocation is determined, the physical infrastructure provider creates
virtual machines, establishes the links between them and hands control over to
the customer (by providing console access to the virtual nodes for instance).

At the heart of the prototype lie MySQL databases where each provider keeps all
information about its substrate topology and embedded virtual networks.  The
database holds a representation of the substrate topology (the \emph{Underlay
(UL) graph}, and for each embedded virtual network there is an \emph{Overlay (OL) graph}
describing the virtual network topology and a \emph{Mapping Layer (ML) graph}
describing the locations of the virtual network's components on the substrate
hosting them.




\subsection{The Embedding Plugin}\label{ssec:mip}


In order to embed CloudNets on the given substrate topology, we pursue a mathematical programming approach and use a \emph{Mixed Integer Program} plugin (short \emph{MIP}). Currently, this is the only implemented embedding plugin in the prototype, but we are working on a fast alternative for a first embedding stage. In order to focus on the technical aspects, we will not describe the used MIP in full detail, but refer the reader to~\cite{grsch-tr}.

The mathematical programming approach is attractive as it is very general and allows us to specify many different types of embeddings and embedding constraints. Especially embeddings on arbitrary Internet infrastructures with heterogeneous resources and links requires a high flexibility. Our Mixed Integer Program supports a simple replacement of the optimization function, without the need to redesign the embedding algorithm. (Rather, the state-of-the-art and optimized algorithms and heuristics provided by standard solvers can be used.)
For instance, one possible objective function may arise in
scenarios where a CloudNet should be embedded in such a manner that the
maximal \emph{load} or \emph{congestion} is minimized. In other scenarios, the
CloudNets should be embedded in a compact manner in order to be
able to shut down the other parts of the physical network, e.g., to
save \emph{energy}.
Moreover, our MIP plugin also supports the migration of (parts of) a CloudNet:
As the CloudNet requests arriving over time may be hard to
predict, certain embeddings computed online in the past may become
suboptimal, and re-embeddings and migrations are necessary.

By computing the embedding that would result
after a migration together with the migration cost, the MIP allows us to determine, e.g., the
\emph{cost-benefit tradeoff} of migration. For example, our algorithm allows to answer
questions such as: Can we migrate CloudNets to a more compact form
such that 20\% of the currently used resources are freed up, and
what would be the corresponding migration cost? It is then left to the
(potentially automated) administrator to decide whether the changes
are worthwhile.

\section{PIP and VNP Roles}

In the following, we will focus on the
\emph{PIP (Physical Infrastructure Provider)} role and the \emph{VNP (Virtual
Network Provider)} role, and will describe their implementation.
The VNO (Virtual Network Operator) and SP (Service Provider)
are not implemented in our prototype yet. 

\subsection{Graph Serialization}

To permit interactions between PIP and VNP roles, a first and important task to be solved is
\emph{graph serialization}.  Since the management nodes for the VNP and the
PIPs in its substrate graph all run on dedicated machines with dedicated
databases, graphs representing the CloudNet topologies have to be serialized
into a database independent representation to be sent across the network. To
this end we developed a \emph{YAML} representation of the resource description
language, along with serialization/deserialization mechanisms to convert the
database representation of a graph into a format suitable for transmission and
back into a database representation on the other side.

\subsection{The PIP}

The PIP role supports arbitrary topology specification and topology
modification including live migration of running nodes (within a PIP's
substrate). It mainly consists of a \emph{Mapping Component} that handles
the Mixed Integer Program formulation, the result interpretation and Mapping Layer
graph generation, and an extensible (plugin based) \emph{Embedding Component}
that embeds a (partial) CloudNet topology on the PIP's substrate given its
overlay and mapping layer graphs. Figure~\ref{fig:architecture} (\emph{left}) gives an
overview of these components.

\subsubsection{Substrate}

The PIP's substrate is largely virtualized in our testbed in Berlin: most substrate
nodes are KVM virtual machines running on the testbed's Sun X4150 machines.
The X4150s are interconnected by an \texttt{IEEE
802.1q}-capable
switch. The virtualized substrate nodes are assigned unique Ethernet addresses
and connected via physical Ethernet interfaces. The CloudNet VNodes are then
created on aforemenioned KVM machines. Each virtual link is assigned a unique 
VLAN tag (picked by the PIP from its assigned range of VLAN tags). These VLAN
tags are set on all data plane interfaces (or switch ports) of the substrate
nodes on the virtual link.

In addition, some of the X4150 machines are configured to serve as
substrate nodes themselves, on top of hosting virtualized substrate nodes. They
can host fully virtualized KVM VNodes (as opposed to the virtualized substrate
nodes which can only host paravirtualized Xen SNodes).

All substrate nodes store their VM images on an NFS share exported by their
controlling PIP. PIPs are responsible for creating VM images and hypervisor
configurations, while the substrate nodes' provisioning scripts take care of
the rest.

Interconnections between links that do not involve VNodes are realized by
creating \emph{bridge} devices on dedicated \texttt{Tunnel Bridge (TBR)}
nodes. Unlike regular substrate nodes, these do not run OpenVSwitch but the
regular Linux bridge. They can host virtual switches and
terminate OpenVPN tunnels.

In particular these bridge devices are used to
establish VLinks spanning across PIP boundaries (\emph{transit links}):
They interconnect the outside segment of these links (within the specifying
VNP's management scope) with the internal segment of these links (within the
PIP's management scope). The mechanism behind this is based on OpenVPN TAP
style layer two tunnels: These are VPN tunnels terminating in virtual Ethernet
devices on both tunnel bridges involved in the transit link. Said virtual
interfaces trunk all VLinks embedded on the transit link in question and are
bridged to the tunnel bridges' data plane interface.

\subsubsection{Cloud Operating System}

The prototype provides similar services as modern Cloud Operating Systems:

\textbf{Automated VM Provisioning.}
The Embedding Component automatically creates VM images for VNodes (typically
by making copies of a template VM image suitable for the hosting substrate
node's hypervisor) and hypervisor configuration files. With everything in place
these VMs can be started through an XMLRPC interface to the hosting substrate
node's provisioning scripts. This interface offers a range of basic VM
provisioning functionalities such as \emph{start}, \emph{stop}, and
\emph{powercycle}.

\textbf{VM Image Caching.\label{clipd}}
Since the I/O performance of the X4150 machines is rather poor and there is
plenty of storage space available we developed a caching scheme to speed up
CloudNet embedding. There is a daemon, \emph{clipd}, that maintains a cache of
VM images on every PIP's exported NFS share (on disk). The Embedding component can
request VM images from this daemon's cache, which are then be moved to their
final location on the NFS share, thus greatly speeding up the whole embedding
process. The \emph{clipd} cache is replenished by a cron job while the Mapping
Layer/Embedding is inactive.

\textbf{Customer Console Access to VNodes.}
Since placement of VNodes is rather opaque to a PIP's customers---and not
relevant as long as their requirements are met---there is a mechanism to
automatically look up any given VNode's hosting substrate node, and offer a
proxied serial console session on that very VNode to a customer requesting
access.

\textbf{Automated Link Provisioning.}
Virtual links are provisioned and brought up automatically. In the Mapping
Layer graph they are split up into segments, one for each substrate network
element they are mapped to (for instance the two substrate nodes on either end
and the switch port the substrate nodes' data plane interfaces are connected
to). For each of these components appropriate Embedding Plugin is called. To
date there are Embedding Plugins handling OpenVSwitch based hosts, traditional
Linux Bridge based hosts and Cisco 4500 series switches (the latter are
configured via SNMP).

\subsubsection{Negotiation and Provisioning Interface}

A PIP controls its substrate nodes and (physical) switch through an
\emph{XMLRPC} configuration interface. The PIP in turn
offers two interfaces to the VNP role: The \emph{Negotiation Interface} and the
\emph{Provisioning Interface}. Both consist of a range
of \emph{XMLRPC} methods.

Negotiation follows a two-stage protocol: The Negotiation Interface allows
for sending preliminary embedding requests for either topology creation or
modification of previously embedded topologies. Once a VNP is sure it wants to
embed a topology permanently (i.e., after receiving the `okays' from all involved PIPs), it 
will send a confirmation request for the preliminary topology.

The VNodes of an embedded topology can then be started and stopped through the
Provisioning Interface.

\subsection{The VNP}

The VNP's high-level architecture is similar to the PIP's, also comprising a
Mapping Component and an Embedding Component. The Mapping Component is for the
most part identical to the PIP's. The major differences are in the embedding
component: (1) since the VNP's substrate consists of entire PIPs rather than
physical nodes and links, the VNP has its own set of embedding plugins to
interact with PIPs' Negotiation Interfaces; (2) the Embedding Component
contains additional code for the creation of \emph{partial graphs} which can be
mapped to a single PIP.

\subsubsection{Substrate}

The VNP's substrate consists of PIP nodes interconnected by \emph{transit
links}\footnote{In our prototype, these are represented by \emph{OpenVPN} (see
\texttt{http://openvpn.net/}) \texttt{tap} style tunnels between the PIPs'
tunnel bridge nodes.}. Hence the VNP is aware of the topology between PIPs (or
a subset of said topology) but does not know anything about the PIP's internal
topology.

\subsubsection{Negotiation and Provisioning Interface}

A library on the VNP level implements the client side functionality of the
PIP's Negotiation and Provisioning Interface. It is used to send partial
topology graphs to PIPs and embed them.

\section{Life of a CloudNet Request}\label{ssec:life}

In order to illustrate the operation of our prototype, we describe the
processing stages a CloudNet topology request graph undergoes from initial
submission to the VNP to final embedding of all its nodes on one or multiple PIPs.
See Figure~\ref{fig:architecture} (\emph{right}) for a graphical outline of these stages.

\subsection{Incoming CloudNet Request}

The CloudNet request in the form of an `overlay graph' (henceforth referred to
as \emph{OL0 graph}) is submitted to the VNP (as a serialized topology graph,
through an XMLRPC interface similar to the PIP's Negotiation Interface).
Subsequently, the VNP's mapping process is started.

\subsection{Substrate Synchronization}

The first step in mapping a CloudNet request is substrate synchronization: The VNP
uses the Negotiation Interface to update the available resources on the PIPs constituting its 
substrate. In order to both protect PIP's business secrets (namely the details of its substrate topology), and give the VNP a rough
approximation of a given PIP's available resources, we simply sum up all the
resources of a given type in the PIP's substrate (i.e., a PIP with three substrate nodes that have
4GB RAM each will report an aggregate resource of 12GB RAM to the VNP). These aggregate resources will then be assigned to the \texttt{/node/host/pip} NE representing this PIP in the VNP's substrate graph. 


\subsection{Solution of VNP-level MIP}

Before proceeding with the mapping process, the VNP completes the OL0
graph, yielding an \emph{OL1 graph}. This is to account for any vagueness on
the upstream entity's part: e.g., values are chosen for those required resources
and features that were left unspecified. For instance the virtualization
technology might not have been specified, allowing the VNP to make a choice of
its own or pass the the vague graph on to downstream players, who will in turn
have the freedom of choice (see the discussion at the end of Section~\ref{sec:flerd}). Next it formulates the Mixed Integer Program (MIP)
for mapping the incoming CloudNet request to its substrate of PIPs. The MIP is
based upon the following variables:
(1) the incoming OL0 graph's resource and feature requirements and constraints,
(2) the available resources and features on the PIPs constituting the VNP's substrate (as far as they are known to the VNP), and
(3) the available resources and features on the transit links between PIPs.

Once the MIP has been solved the result is translated back into a mapping layer
graph, mapping all of the OL1 graph's Network Elements (NEs) to PIPs and transit
links.

\subsection{Partial Graph Generation}

Now the VNP iterates through the OL1 graph and creates a set of partial graphs,
one for each PIP. Each of these partial graphs consists of all the NEs mapped
to the PIP in question.

In the course of partial graph generation, stubs for transit links are created.
These consist of two special network element types: \texttt{/node/host/pip} and
\texttt{/link/transit}.

The \texttt{/link/transit} \emph{NE} specifies the VLAN to use for this transit link
(we opted for letting the VNP pick transit VLANs to avoid implementing an error
prone PIP-to-PIP negotiation mechanism for this purpose). This closely matches
the most likely real-world scenario, too: A VNP might have an embedding plugin for a
transit provider (for instance someone running an MPLS network) with black boxes
located at both PIPs' network edges. This embedding plugin requests a transit link
between these two PIPs and receives some kind of link identifier (such as a
VLAN tag). The VNP then communicates this link identifier to the PIPs on either
end, which allows both PIPs to connect the black box to their local segment of
the virtual link.

\subsection{Serialization and Transmission}

Once the partial graphs have been assembled, each of them is serialized and
sent to its hosting PIP through the Negotiation Interface. This currently takes
place in a serial manner in our prototype, i.e., partial graphs are sent to PIPs one after the
other. If one of the PIPs in the chain turns out to be unable to embed its
partial graph the VNP rolls back the already embedded partial graphs (deletes
them) and reports a failure status to the upstream entity that sent the
original CloudNet request.

\subsection{Formulation and Solution}

Upon receipt of a partial graph a PIP will---much like the VNP---complete the
OL0 graphs to yield OL1 graphs, and formulate the MIP for mapping the incoming
partial CloudNet request to its substrate of physical hardware. The program is
based on the following variables:
(1) the incoming partial graph's resource and feature requirements and constraints,
(2) the available resources and features on this PIP, and
(3) the available resources and features on this PIP's outgoing transit links leading to requested destinations.

Once the MIP has been solved the result is translated back into a mapping layer
graph, mapping all of the OL1 graph's Network Elements to substrate links
and nodes. Control passes to the Embedding Component now which will configure
the topology's VLANs on the switch as specified by the Mapping Component and
create the VNodes' images and hypervisor configuration. With everything in
place the PIP reports success to the VNP which sent the (partial) CloudNet creation
request.

\subsection{Bootup and Handoff}

Once success has been reported back to the VNP by all PIPs involved in the CloudNet
creation request, the VNP confirms all these requests through the PIPs' Negotiation Interface, uses the PIPs' Provisioning Interface to 
start the newly created VNodes (the links will already have been brought up in
the course of negotiation). Finally, the VNP hands control over the CloudNet to the
requesting upstream entity (typically in the form of console interfaces).

\section{Outlook}

 Our proof-of-concept prototype is optimized for flexibility and generality rather than performance.
 Although it is already fully functional, several extensions are planned.

Currently, our prototype only supports migrations within a PIP. To enable VNode
migration across PIP boundaries, a mechanism is required that orchestrates
migration on the VNP side and makes sure space for migrating VNodes is
allocated and ready before migration. This implies slight modifications to the
negotiation and provisioning interfaces and some changes to the PIP's embedding
component.

Moreover, negotiations with PIPs
currently take
place in a serial manner. Yet it would be desirable to negotiate with all
involved PIPs in parallel, thus exploiting the distributed nature of the
architecture.
Finally, we will also need to address the problem of \emph{terminal attachment}
and develop a mechanism for attaching clients to CloudNets. We are currently
envisioning a role that presents itself as a PIP on the CloudNet side with a
special \emph{customers} resource that contains the number of currently
connected terminals. Thus, scenarios such as a Service Provider specifying
CloudNet requirements like ``My service must reach 20k or more terminals!'' or
``My service must reach all Swiss users at a latency below 20ms!'' are
supported.


\begin{thebibliography}{9}
\bibitem{netperf}
G.~Wang and E.~Ng.
\newblock The impact of virtualization on network performance of Amazon EC2 data center.
\newblock In {\em Proc. IEEE INFOCOM}, 2010.

\bibitem{d3}
C.~Wilson, H.~Ballani, T.~Karagiannis, and A.~Rowtron.
\newblock Better never than late: meeting deadlines in datacenter networks.
\newblock In {\em Proc. ACM SIGCOMM Conference}, pages 50--61, 2011.

\bibitem{j-sdn}
D.~Drutskoy, E.~Keller, and J.~Rexford.
\newblock Scalable network virtualization in software-defined networks.
\newblock In {\em Proc. IEEE Internet Computing}, 2012.

\bibitem{visa09virtu}
G.~Schaffrath, C.~Werle, P.~Papadimitriou, A.~Feldmann, R.~Bless,
  A.~Greenhalgh, A.~Wundsam, M.~Kind, O.~Maennel, and L.~Mathy.
\newblock Network virtualization architecture: Proposal and initial prototype.
\newblock In {\em Proc. ACM SIGCOMM VISA}, pages 63--72, 2009.

\bibitem{icccn12stefan}
G.~Schaffrath, S.~Schmid, I.~Vaishnavi, A.~Khan, and A.~Feldmann.
\newblock A resource description language with vagueness support for
  multi-provider cloud networks.
\newblock In {\em Proceedings of International Conference on Computer
  Communication Networks (ICCCN)}, 2012.


\bibitem{grsch-tr}
G.~Schaffrath, S.~Schmid, and A.~Feldmann.
\newblock Optimizing long-lived cloudnets with migrations.
\newblock In {\em Proc. 5th IEEE/ACM International Conference on Utility and
  Cloud Computing (UCC)}, 2012.



\vspace{4mm}

\end{thebibliography}
\end{document}